# Improved Nondestructive Isotopic Analysis with Practical Microcalorimeter Gamma Spectrometers
LA-UR-21-23156

[1]Mark Croce, [2]Daniel Becker, [1]Katrina E. Koehler, [3]Joel Ullom

[1]Los Alamos National Laboratory, Los Alamos, New Mexico, USA
[2]University of Colorado, Boulder, Colorado, USA
[3]National Institute of Standards and Technology, Boulder, Colorado, USA

## Abstract

Advances in both instrumentation and data analysis software are now enabling the first ultra-high-resolution microcalorimeter gamma spectrometers designed for implementation in nuclear facilities and analytical laboratories. With approximately ten times better energy resolution than high-purity germanium detectors, these instruments can overcome important uncertainty limits. Microcalorimeter gamma spectroscopy is intended to provide nondestructive isotopic analysis capabilities with sufficient precision and accuracy to reduce the need for sampling, chemical separations, and mass spectrometry to meet safeguards and security goals. Key milestones were the development of the SOFIA instrument (Spectrometer Optimized for Facility Integrated Applications) and the SAPPY software (Spectral Analysis Program in PYthon). SOFIA is a compact instrument that combines advances in large multiplexed transition-edge sensor arrays with optimized cryogenic performance to overcome many practical limitations of previous systems. With a 256-pixel multiplexed detector array capable of 5000 counts per second, measurement time can be comparable to high-purity germanium detectors. SAPPY was developed to determine isotopic ratios in data from SOFIA and other microcalorimeter instruments with an approach similar to the widely-used FRAM software. SAPPY provides a flexible framework with rigorous uncertainty analysis for both microcalorimeter and HPGe data, allowing direct comparison. We present current results from the SOFIA instrument, preliminary isotopic analysis using SAPPY, and describe how the technology is being used to explore uncertainty limits of nondestructive isotopic characterization, inform safeguards models, and extract improved nuclear data including gamma-ray branching ratios.

## 1. Introduction

Ultra-high resolution microcalorimeter technology has advanced rapidly in the past several years and is under evaluation as a next-generation nondestructive isotopic analysis tool for safeguards and security applications [1-5]. The first gamma spectrometers designed for implementation in nuclear facilities and analytical laboratories are now being built and deployed. WIth approximately ten times better energy resolution than high-purity germanium detectors, these instruments can overcome important uncertainty limits in nondestructive assay and reduce the need for destructive analysis based on sampling, chemical separations, and mass spectrometry. This paper reviews the current status of microcalorimeter gamma spectroscopy instrumentation and analysis software, and how recent results are being used to explore uncertainty limits of nondestructive isotopic characterization, inform safeguards models, and extract improved nuclear and atomic data including gamma-ray branching ratios and X-ray line widths needed for isotopic analysis.

## 2. SOFIA Microcalorimeter Spectrometer

SOFIA (Spectrometer Optimized for Facility Integrated Applications) is a compact instrument that combines advances in large multiplexed transition-edge sensor arrays with optimized cryogenic performance to overcome many practical limitations of previous systems [3-5]. Figure 1 shows the main components of the instrument. A primary goal of the instrument was to reduce the infrastructure required for operation, in order to enable simplified deployment in a facility. Whereas previous microcalorimeter systems typically use large cryostats requiring 3-phase power and cooling water, SOFIA is built around a compact tabletop cryostat with a low-power, air-cooled pulse tube cryocooler. Only single-phase 220V power (~3 kW) is required for the cryocooler, which is comparable to a large window air conditioner and compatible with nearly any facility worldwide. No liquid cryogens are present in the system. An adiabatic demagnetization refrigerator, which is much easier to operate than a dilution refrigerator and requires no $^3$He, provides cooling to the detector operating temperature of 90 mK for up to 48 hours at a time following an automated 2-hour regeneration cycle.

With a 256-pixel multiplexed detector array capable of 5000 counts per second, measurement time can be comparable to high-purity germanium detectors. Frequency-division multiplexed readout makes extensive use of commercial microwave components and implements much of the complexity in FPGA firmware [6]. Software for data processing automates the task of combining individual pixel responses into a single spectrum [7]. As with a germanium detector, samples are placed next to the instrument for measurement. Samples can be in nearly any form, size, or container type as long as there is sufficient gamma-ray activity in the 30-300 keV range for which the detectors are optimized. Energy resolution as good as 59 eV FWHM at 97 keV has been demonstrated. For high-rate Pu measurements at 15-20 counts per second per pixel, energy resolution of 70-90 eV FWHM is typical. A second instrument based on the same high-throughput multiplexed architecture is being built for permanent installation in an analytical laboratory in 2021.

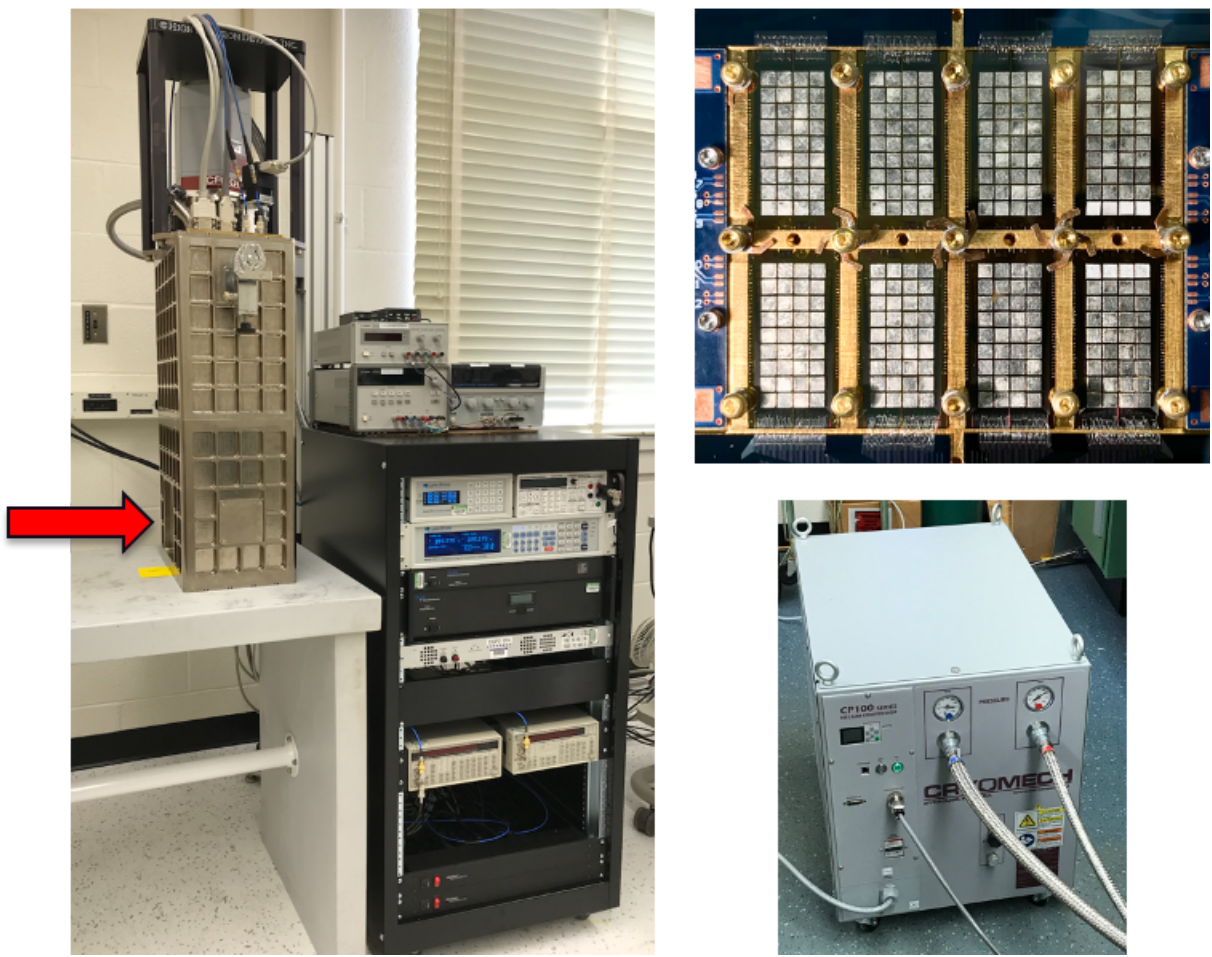

**Figure 1:** The compact SOFIA microcalorimeter spectrometer consists of a cryostat (left), 256-pixel multiplexed transition-edge sensor microcalorimeter array (top right) and an air-cooled compressor (bottom right). Items to be measured are placed next to the cryostat in front of the side-looking window (indicated by red arrow), or the instrument can be moved next to a glove box or hot cell port.

## 3. SAPPY Quantitative Analysis Software

To analyze microcalorimeter spectra, software called SAPPY (Spectral Analysis Program in PYthon) [5] was developed to determine isotopic ratios in an approach similar to the widely-used FRAM software [8-9]. SAPPY provides a flexible analysis framework with rigorous uncertainty analysis for both microcalorimeter and HPGe data, allowing direct comparison. Compared to FRAM, SAPPY offers a more detailed uncertainty analysis including the option to include uncertainty from gamma-ray energies or X-ray line widths. Flexible peak fitting options including different background models, peak and tail shapes, and options to fix the relative positions of peaks can be applied to a wide range of measured spectra. Figure 2 shows an example output of SAPPY for a Pu reference material containing approximately 14 wt% $^{240}$Pu. With minimal additional development, SAPPY can be extended to uranium enrichment measurements.

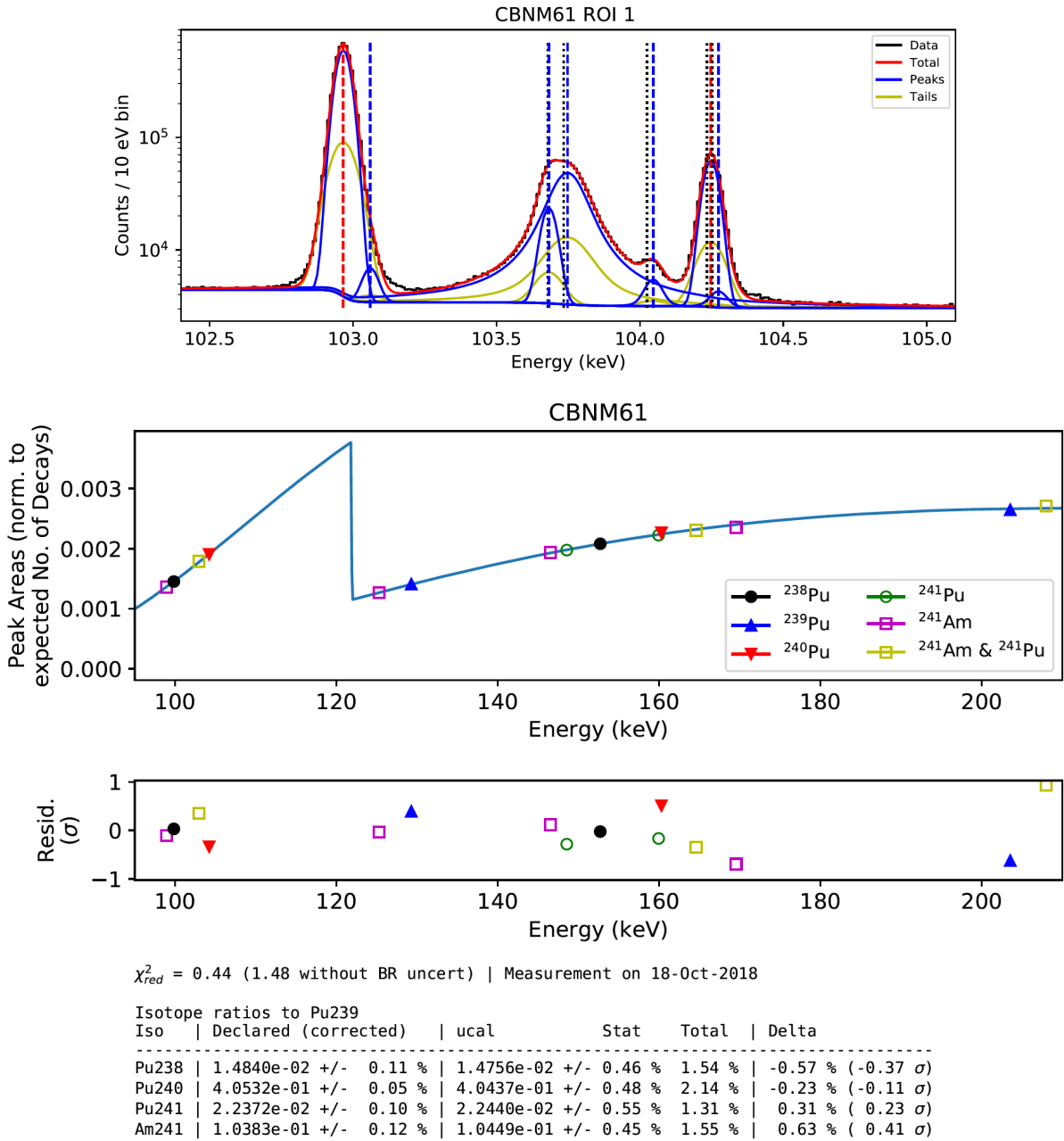

**Figure 2:** Example output of SAPPY for a Pu reference material spectrum with multi-peak fit in one spectral region (top), determined relative efficiency curve (middle), and efficiency curve residuals (bottom). The energy resolution of the microcalorimeter data is especially valuable in the Pu X-ray region near 100 keV. Note the separation of the 104.2 keV peak from $^{240}$Pu. [5]

## 4. Results

SOFIA is now being used to evaluate the utility and performance of ultra-high-resolution microcalorimeter gamma spectroscopy for safeguards and security purposes throughout the nuclear fuel cycle. Potential applications include uranium enrichment determination in fresh fuel, spent fuel characterization, determination of actinides in aqueous and electrochemical spent fuel separation processes, isotopic analysis of purified Pu, U, and U/TRU products, and even liquid fuel in advanced reactors [4-5]. Laboratory measurements of samples representing each of these categories are in progress, and early results show significant potential for microcalorimetry to improve nondestructive assay performance in such applications.

Uranium enrichment measurements benefit from improved energy resolution in the 90-94 keV energy range, where $^{234}$Th gamma rays and Th X-rays can be used to determine the ratio of $^{235}$U/$^{238}$U in material approximately 6 months or more after chemical purification. However, microcalorimeter energy resolution may enable measurements that are not feasible with traditional nondestructive assay technology. One such example is shown in Figure 3. Direct quantification of $^{238}$U in freshly-purified uranium material may be possible through use of the 113.5 keV gamma ray from the direct decay of $^{238}$U. This peak is clearly observed in microcalorimeter spectra from low-enriched or natural uranium reference materials consisting of 200-230 g of $U_3O_8$ each in aluminum cans. Samples optimized to reduce Compton scattering background may enable accurate measurements at higher enrichment levels.

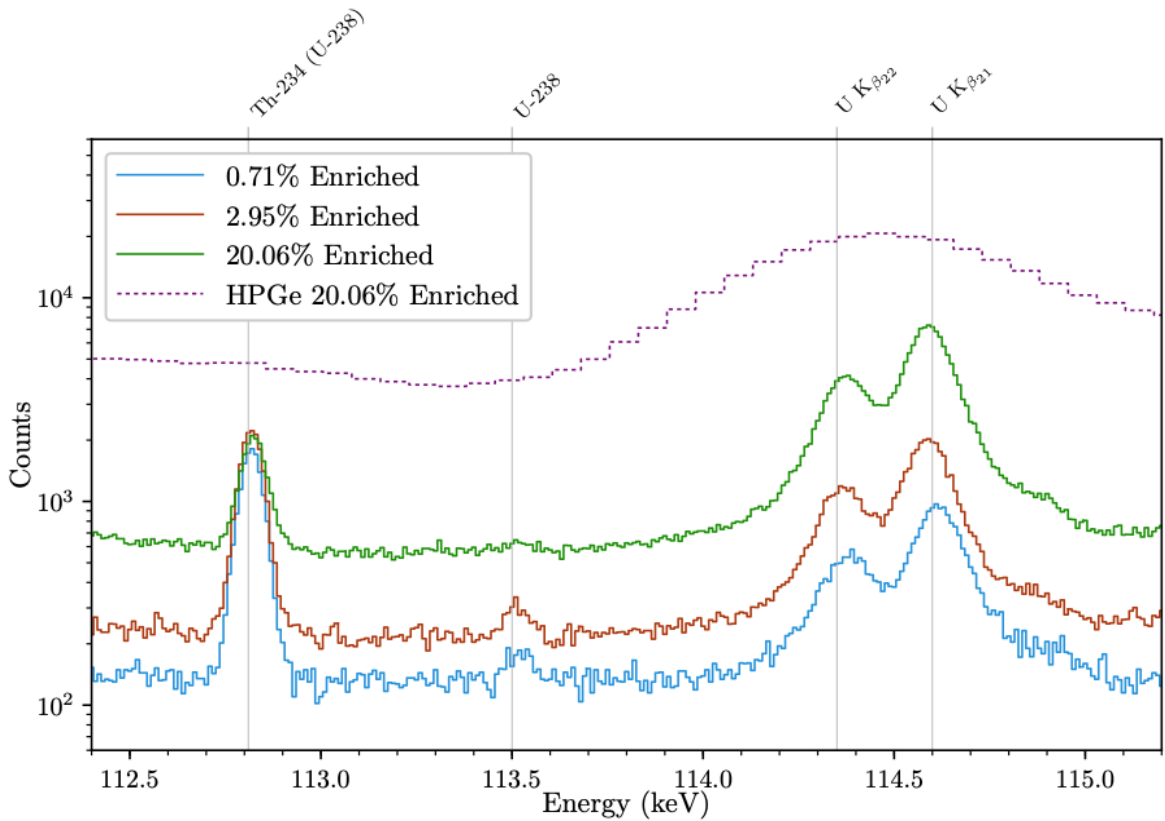

**Figure 3:** Observation of the 113.5 keV peak from the direct decay of $^{238}$U, usually considered too weak to measure, suggests that microcalorimeter spectroscopy may provide a robust method of verifying uranium enrichment in freshly purified material where Th decay products have not yet reached secular equilibrium with $^{238}$U [4].

In an analogous problem for isotopic analysis of Pu materials, $^{237}$Np is typically quantified using gamma ray peaks from its decay product $^{233}$Pa. In Figure 4, the 86.5 keV gamma ray from direct $^{237}$Np decay is observed in the microcalorimeter spectra for two $PuO_2$ reference materials packaged in stainless steel containers. The 86.6 keV peak from $^{233}$Pa can also be resolved. The ratio of these two peaks could be useful to determine the chemical age of Pu materials, especially as they are extremely close in energy and therefore the uncertainty component from knowledge of the efficiency curve will be minimal.

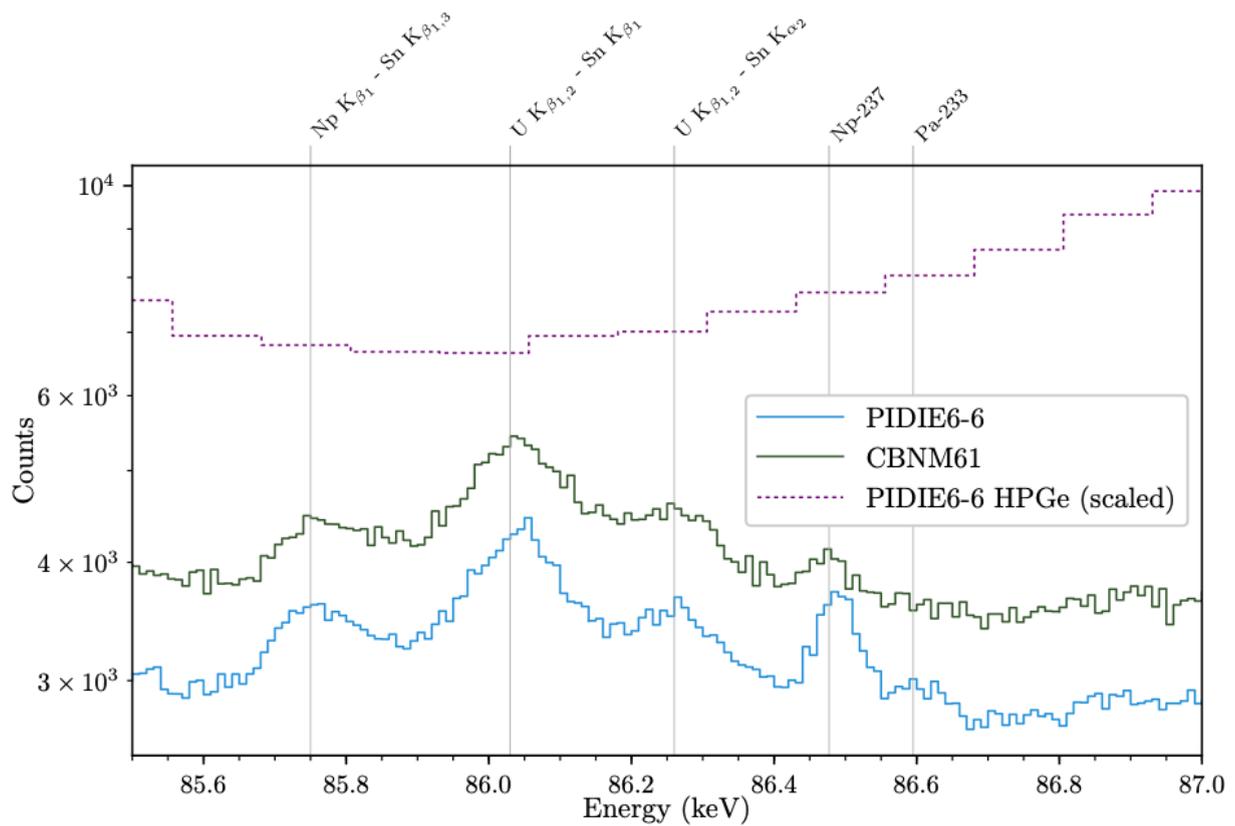

**Figure 4:** Observation of the 86.5 keV peak from the direct decay of [237]Np, usually considered too weak to measure, suggests that microcalorimeter spectroscopy may provide a robust method of quantifying this nuclide in Pu materials. Note that the 86.6 keV gamma ray from [233]Pa is also resolved in the PIDIE6-6 spectrum [4].

It has been shown that for reference material spectra with sufficient counting statistics, the uncertainty in determined isotopic ratios in the tabulated branching ratios, X-ray line widths, and gamma-ray energies. As a result, quantitative analysis methods can be used "in reverse" with well-characterized samples to determine key nuclear data for non-destructive isotopic analysis. In particular, improved branching ratios were determined for Pu and [241]Am in the 125-208 keV region, which have not been measured since the mid-1990s [10]. Table 1 shows selected branching ratios where significant improvements in uncertainties were obtained relative to National Nuclear Data Center (NNDC) values [11-12]. Further improvements in branching ratio uncertainty are possible with longer measurement times for increased counting statistics, and work is underway to determine branching ratios for gamma ray energies below the Pu K-edge. These results demonstrate the potential of microcalorimeter measurements to improve the performance of HPGe-based isotopic analysis through the use of more precise nuclear data.

Table 1: Selected gamma ray branching ratios determined through analysis of microcalorimeter spectra. Uncertainty in % relative standard deviation is given in parentheses.

| Energy (keV) | Nuclide | NNDC BR [11-12] | Microcal-determined BR [10] |
|---|---|---|---|
| 125.3 | $^{241}$Am | 4.08e-3 (2.5%) | 4.08e-3 (1.0%) |
| 144.20 | $^{239}$Pu | 2.83e-4 (2.1%) | 2.87e-4 (1.0%) |
| 146.09 | $^{239}$Pu | 1.19e-4 (2.5%) | 1.22e-4 (1.4%) |
| 146.55 | $^{241}$Am | 4.61e-4 (2.6%) | 4.75e-4 (0.75%) |
| 150.04 | $^{241}$Am | 7.40e-5 (3.0%) | 7.76e-5 (1.3%) |

The high energy resolution of microcalorimeter detectors combined with the high statistics measurements of the multiplexed system made it possible to identify and quantify two new gamma transitions at 207.8 keV and 208.2 keV from $^{241}$Am near the 208.0 keV peak. These transitions were compared with nuclear levels in $^{237}$Np and potential transition candidates were identified. Using measurements of certified Pu items as well as pure $^{241}$Am sources, the branching ratios of these satellite peaks were quantified using the ratio of the counts within the satellite peaks to the counts in the 208.0 keV peak that came from $^{241}$Am [13].

The SAPPY code is currently being used to evaluate uncertainty limits in nondestructive plutonium isotopic analysis achievable with microcalorimeter and HPGe detectors. Preliminary comparison results in Table 2 are expressed as the ratio of the microcalorimeter difference from the declared value to the HPGe difference from the declared value. Numbers less than 1 indicate better results for microcalorimetry and numbers greater than 1 would indicate better results for HPGe. Cells shaded green indicate a substantial advantage for microcalorimetry, and no shading indicates comparable performance for both detector types. It is worth noting that these results are preliminary. PIDIE 6-6 contained 0.5 g of Pu oxide (25 wt% $^{240}$Pu) and the microcalorimeter spectrum had 99 million counts, and CBNM61 contained 6.6 g of Pu oxide (27 wt% $^{240}$Pu) and the microcalorimeter spectrum had 64 million counts. A separate publication with a detailed quantitative comparison between HPGe and microcalorimeter is in preparation by the authors. These results, and similar analysis for materials encountered in advanced fuel cycles, are an important input to facility safeguards models [5].

Table 2: Preliminary Comparison of Pu Isotopic Analysis Performance with Microcalorimeter and High-Purity Germanium Detectors

| Isotope Ratio | CBNM61 | PIDIE6-6 |
|---|---|---|
| $^{238}$Pu/$^{239}$Pu | 0.05 | 0.13 |
| $^{240}$Pu/$^{239}$Pu | 0.86 | 0.70 |
| $^{241}$Pu/$^{239}$Pu | 0.02 | 0.44 |
| $^{241}$Am/$^{239}$Pu | 0.15 | 0.99 |

## 5. Summary

As a result of extensive technology development over the past several years, ultra-high resolution microcalorimeter gamma spectroscopy is now offering real capabilities for nondestructive isotopic analysis throughout the nuclear fuel cycle. The first instruments such as SOFIA are in the process of being evaluated in nuclear facilities and analytical laboratories. The advanced isotopic analysis code SAPPY is being used to test uncertainty limits of nondestructive assay and determine improved nuclear data needed for safeguards. As the use of microcalorimeter technology continues to advance, it is anticipated to be a valuable tool for cost-effective and robust material accountancy in nuclear safeguards.

## 6. Acknowledgements

This work is supported by the United States Department of Energy, Office of Nuclear Energy, Material Protection Accounting and Control Technologies (MPACT) program and Nuclear Energy University Program (NEUP). The microcalorimeter development team includes Katrina Koehler, Matthew Carpenter, Katherine Schreiber, Chandler Smith, Daniel McNeel, and Sophie Weidenbenner of Los Alamos National Laboratory; Nathan Ortiz, Johnathan Gard, and Abigail Wessels of the University of Colorado; and Douglas Bennett, Gene Hilton, John A.B. Mates, Carl Reintsema, Daniel Schmidt, Daniel Swetz, and Leila Vale of the National Institute of Standards and Technology.

## 7. Keywords

Nondestructive assay, isotopic analysis, microcalorimeter, nuclear safeguards

## 8. Author Biographies

Mark Croce is a Scientist in the Safeguards Science and Technology group in the Nuclear Engineering and Nonproliferation division of Los Alamos National Laboratory. He has over 15 years of experience in the development of cryogenic detector systems and novel instrumentation for nondestructive nuclear material analysis. He holds a bachelor's degree in physics from the University of California, Santa Barbara.

Daniel Becker is a Research Associate at the University of Colorado and a member of the NIST Quantum Sensors Group. He has 14 years of experience working with cryogenic detector systems, covering both the development of new scientific instruments and new data analysis techniques and algorithms. He holds a PhD in Physics from the University of Colorado.

Katrina E. Koehler graduated with her PhD and MA in nuclear physics from Western Michigan University and her BS in Mathematics and Physics from Houghton College. She is a research scientist at Los Alamos National Laboratory, where she works in the Nuclear Safeguards Science and Technology Group. Her research focuses on the use of low temperature detectors for safeguards science and basic nuclear physics as well as developing neutron multiplicity algorithms and simulation of neutron multiplicity data.

Joel Ullom is a Lecturer in the Department of Physics at the University of Colorado Boulder and the leader of the NIST Quantum Sensors Group. He has worked in the field of cryogenic electronics since 1994 and has led the development of spectrometers for materials analysis based on multiplexed arrays of transition-edge sensors. He holds a PhD in Physics from Harvard University.